\RequirePackage[2020-02-02]{latexrelease}
\documentclass[aps,pra,showpacs,twocolumn,superscriptaddress]{revtex4}
\usepackage{array}
\usepackage{booktabs}
\usepackage{tabu}
\usepackage{dcolumn}
\usepackage{amsmath}
\usepackage{amsfonts}
\usepackage{float}
\usepackage{amssymb}
\usepackage{graphicx,color}
\usepackage[colorlinks={true}]{hyperref}
\hypersetup{colorlinks=true,linkcolor=red,citecolor=blue,urlcolor=blue}
\usepackage{graphicx}
\usepackage{subfigure}
\usepackage{graphicx}
\usepackage{dcolumn}
\usepackage{bm}

\def\be{\begin{equation}}
  \def\ee{\end{equation}}
\def\bea{\begin{eqnarray}}
\def\eea{\end{eqnarray}}
\def\f{\frac}
\def\n{\nonumber}
\def\l{\label}
\def\p{\phi}
\def\o{\over}
\def\R{\rho}
\def\pa{\partial}
\def\om{\omega}
\def\na{\nabla}
\def\P{\Phi}
\begin{document}
\title{Quantum dynamical speedup for correlated initial states}
\author{Alireza Gholizadeh}
\affiliation{Faculty of Physics, Urmia University of Technology, Urmia, Iran.}
\author{Maryam Hadipour}
\affiliation{Faculty of Physics, Urmia University of Technology, Urmia, Iran.}
\author{Soroush Haseli}
\email{soroush.haseli@uut.ac.ir}
\affiliation{Faculty of Physics, Urmia University of Technology, Urmia, Iran.}
\affiliation{School of Physics, Institute for Research in Fundamental Sciences (IPM), P.O. Box 19395-5531, Tehran, Iran.}
\author{Saeed Haddadi}
\affiliation{School of Physics, Institute for Research in Fundamental Sciences (IPM), P.O. Box 19395-5531, Tehran, Iran.}
\affiliation{Saeed's Quantum Information Group, P.O. Box 19395-0560, Tehran, Iran}
\author{Hazhir Dolatkhah}
\affiliation{RCQI, Institute of Physics, Slovak Academy of Sciences, D\'{u}bravsk\'{a} cesta 9, 84511 Bratislava, Slovakia}
\date{\today}
\def\be{\begin{equation}}
  \def\ee{\end{equation}}
\def\bea{\begin{eqnarray}}
\def\eea{\end{eqnarray}}
\def\f{\frac}
\def\n{\nonumber}
\def\l{\label}
\def\p{\phi}
\def\o{\over}
\def\R{\rho}
\def\pa{\partial}
\def\om{\omega}
\def\na{\nabla}
\def\P{\Phi}

\begin{abstract}
The maximal evolution speed of any quantum system can be expressed by the quantum speed limit time. In this paper, we consider a model in which the system has a correlation with the environment. The influence of the initial correlation between the system and environment on the quantum speed limit is investigated. It is shown that the appearance of non-Markovianity effects causes the speedup of quantum evolution. Moreover, we demonstrate the dependence of quantum dynamical speedup on the quantum coherence of the correlated initial state.

\end{abstract}
\maketitle

\section{INTRODUCTION}	
The question of how fast a quantum system can transform from an initial state to an orthogonal state is a starting point to  study the quantum speed limit (QSL). The main goal of this study  is to find the general bounds that exist for any system which can be used to limit the time (from below) takes for the state of a system to be distinguishable from its initial state. The bounds known as QSL time. Indeed, QSL time is the shortest possible time for the evolution of the system from an initial state to the final orthogonal state. The study of QSL time has particular importance for quantum
communication \cite{Bekenstein1981}, computation \cite{Lloyd2000}, metrology \cite{Giovannetti2011}, and many other areas of quantum physics. The QSL time can  be used to obtain the shortest time needed to charge the quantum battery \cite{Campaioli20188} and it is also used to find the minimum time required to implement quantum gates in quantum computing \cite{Ashhab2012}. In Ref. \cite{Mohan2021}, using the geometry of the quantum state space, the inverse QSL is introduced and its application in quantum batteries is also discussed.

The first results of the studies on QSL time for closed quantum systems were presented by Mandelstam and Tamm (MT) \cite{Mandelstam1945}. They showed that for an evolution generated by a time-independent Hamiltonian, the shortest possible time for the transformation of an initial pure state to its final orthogonal state is bounded by
\begin{equation}\label{MT}
\tau \geq \frac{\pi \hbar }{2 \Delta E},
\end{equation}
where $\Delta E  =\sqrt{\langle \psi \vert H^{2} \vert \psi \rangle - \langle \psi \vert H \vert \psi\rangle^2} $ is the standard deviation of time-independent Hamiltonian $H$ and $\hbar$ is the reduced Planck constant.   The key point of the  MT bound (\ref{MT}) is its dependence on the standard deviation of the system energy.  A lot of work has been done to extend the MT bound, however, the most effective result has been obtained by Margolus and Levitin (ML), who have presented the new bound as \cite{Margolus1998}
\begin{equation}\label{ML}
\tau \geq \frac{\pi \hbar}{2 E},
\end{equation}
where $E=\langle H \rangle$ is the average energy over the ground state of the system. The bound in Eq. (\ref{ML}) is known as ML bound.  So, for unitary evolution that connects two pure and orthogonal states, the bound for the QSL is not unique, and usually a comprehensive bound can be introduced by combining  MT and  ML bounds as follows
\begin{equation}
\tau \geq \max \lbrace \frac{\pi \hbar }{2 \Delta E}, \frac{\pi \hbar}{2 E} \rbrace.
\end{equation}
In the practical scenario, real quantum systems interact with their surroundings,  such systems are called open quantum systems \cite{Breuer2002,Czerwinski2022}. In recent years, the study of the QSL for open quantum systems has received much attention \cite{Deffner2013,del Campo2013,Taddei2013,Pires2016,Jing2016,Pintos2022,Xu2019,haseli2022}. Geometric approach is usually used to obtain the desired bound for the QSL in open quantum systems. In Ref. \cite{Taddei2013}, Taddei et al. have introduced a bound for the QSL in open quantum systems by using Fisher information for the total Hilbert space of the system and its environment. Escher et al. developed their results in Ref. \cite{Escher2011}. The authors of Ref. \cite{del Campo2013} have used relative purity to introduce a QSL bound  for  open quantum systems. They showed that when the evolution is in the Lindblad form, the bound is equivalent to the MT bound. Moreover,  Deffner
and Lutz \cite{Deffner2013} have introduced a bound for the QSL using the Bures angle, which covers both  MT and ML bounds. They also showed that the non-Markovian effects can speedup the quantum process. In recent years, due to the fact that it is difficult to access an initial pure state in practical scenarios, studying the QSL for mixed initial states has been the subject of some works \citep{Zhang2014,xiong2018}. In addition to providing different bounds, some works have been done on the issue of QSL, such as the dependence of QSL on the initial state \cite{Wu2015}, and many other works \cite{Carabba2022,Xu2016,Dehdashti2015,Zhang2015,Xu2014,Mondal2016,Ektesabi2017,Sun2015,Cai2017,Liu2015,Campaioli2022,
Satoshi2022,Campbell2022,Srishty2022,Kazutaka2022,Brij2022,Kang2022,Niklas2022,Maxwell2022}.

In this paper, a comprehensive bound is considered for the QSL based on the function of relative purity, which can be applied to any mixed initial state \cite{xiong2018}. Based on this bound, we investigate the QSL for a correlated initial state. It is observed that the QSL depends on the non-Markovianity of the evolution, the amount of initial correlation between the system and the environment, and the initial coherence of the system. We show that in addition to the non-Markovian effects, the initial correlation between the system and the environment can speedup the evolution of the quantum system.   Besides, we reveal that strengthening the coupling of the system with the environment enhances the bound on the QSL.

The work is organized as follows. In Sec. \ref{Model}, the main structure of the model that will be used in this work is presented. In Sec. \ref{Non-Markovian}, the non-Markovian feature of the model and  parameters that are effective in the non-Markovian evolution will be studied. In Sec. \ref{QSL}, the QSL for a correlated initial state will be investigated. Finally, the results will be summarized in Sec.\ref{conclusion}.

\section{Model}\label{Model}
Here, we consider a model in which a two-level system $\mathcal{S}$  is coupled to its environment $\mathcal{R}$. In this model, just the pure decoherence of the qubit is considered as a mechanism
for decoherence and the energy dissipation is ignored. The model can be described by the following Hamiltonian \cite{Dajka2010}
\begin{equation}
H=H_\mathcal{S}+H_\mathcal{R} + H_{\mathcal{S R}},
\end{equation}
where $H_\mathcal{S}=\omega_0 \sigma_z$ indicates the Hamiltonian of the system with $\omega_0$ which is the qubit energy splitting, $\quad H_\mathcal{R}=\int_0^{\infty} d \omega h(\omega) a^{\dagger}(\omega) a(\omega)$ denotes the environment Hamiltonian, and
$H_\mathcal{SR}=\int_0^{\infty} d \omega \sigma_z \left[g^*(\omega) a(\omega)+g(\omega) a^{\dagger}(\omega)\right]$ is the interaction Hamiltonian. In the above Hamiltonian, $\sigma_z$ is the $z$-component of Pauli matrix, $a(\omega)$ and $a^\dag(\omega)$ are the bosonic annihilation  and creation operators, respectively, $h(\omega)$ is the real-valued spectrum function which describes the environment, and $g(\omega)$ is the function that characterizes the coupling. The whole Hamiltonian can be rewritten in block-diagonal form as \cite{Spohn1978}
\begin{equation}
H=\operatorname{diag}\left[H_{+}, H_{-}\right], \quad H_{\pm}=H_\mathcal{R} \pm H_\mathcal{SR} \pm \omega_0 \mathbb{I}_\mathcal{R},
\end{equation}
 where $\mathbb{I}_\mathcal{R}$ is identity operator in environment Hilbert space. The correlated initial state for the system-environment can be written as
\begin{equation}\label{ins}
|\Psi(0)\rangle=c_e|e\rangle \otimes\left|\Omega_0\right\rangle+c_g|g\rangle \otimes\left|\Omega_\lambda\right\rangle,
\end{equation}
where $|e\rangle$ and $|g\rangle$ describe the excited and ground states of the system respectively,  $c_{g}$ and $c_{e}$ are two non-zero complex numbers that satisfy $\left|c_g\right|^2+\left|c_e\right|^2=1$. Besides,
$\left|\Omega_0\right\rangle$ and $\left|\Omega_\lambda\right\rangle$ are the states of the environment where $\left|\Omega_0\right\rangle$ denotes an environment ground state and
\begin{equation}\label{en1}
 \left|\Omega_\lambda \right\rangle = \eta_\lambda^{-1}[(1-\lambda)]\left|\Omega_0\right\rangle + \lambda D(f)  \left|\Omega_0\right\rangle],
\end{equation}
where
$D(f)=\exp \left\{\int_0^{\infty} d \omega\left[f(\omega) a^{\dagger}(\omega)-f^*(\omega) a(\omega)\right]\right\}$ is the displacement operator for an arbitrary square-integrable function $f$. Considering that the state presented in Eq.(\ref{en1}) should be normalized $ \langle \Omega_\lambda \vert \Omega_\lambda \rangle =1$, the coefficient $\eta_\lambda$ is obtained as follows
\begin{equation}
\eta_\lambda^2=(1-\lambda)^2+\lambda^2+2 \lambda(1-\lambda) \operatorname{Re}\left\langle\Omega_0 \mid D(f) \mid \Omega_0\right\rangle.
\end{equation}
Above, the term $\operatorname{Re}$ means the real part of a complex number. The parameter $\lambda\in [0, 1]$ specifies the initial entanglement between the system and the environment. If this value is  $\lambda=0$, it means that the system and the environment are initially uncorrelated, while for $\lambda=1$ there exists the greatest possible entanglement between the system and the environment. By considering the initial state of the composite system \eqref{ins}, the state of the system-environment at time $t$ can be written as follows
\begin{equation}
|\Psi(t)\rangle=c_e|e\rangle \otimes\left|\psi_{+}(t)\right\rangle+c_g|g\rangle \otimes\left|\psi_{-}(t)\right\rangle,
\end{equation}
where $\left|\psi_{+}(t)\right\rangle = \exp(-i H_+ t) \vert \Omega_0 \rangle$ and $\left|\psi_{-}(t)\right\rangle = \exp(-i H_- t) \vert \Omega_\lambda \rangle$. The reduced density matrix of the system $\rho_\mathcal{S}^{\lambda}(t)$ is obtained by giving partial trace over environment as
\begin{equation}\label{tr}
\rho_\mathcal{S}^{\lambda}(t)=\operatorname{tr}_\mathcal{R}\left[\vert \Psi(t) \rangle  \langle \Psi(t) \vert \right],
\end{equation}
with more details, the explicit form of the above density matrix is obtained as follows
 \begin{equation}\label{rhot}
\rho_\mathcal{S}^{\lambda}(t)=\left(\begin{array}{cc}
\left|c_{e}\right|^2 & c_{e} c_{g}^* \kappa_\lambda(t) \\
c_{e}^* c_{g} \kappa_\lambda^*(t) & \left|c_{g}\right|^2
\end{array}\right),
 \end{equation}
 with
 \begin{equation}\label{kappat}
\kappa_\lambda(t)=\eta_\lambda^{-1} e^{- i 2  \omega_0 t} e^{-r(t)}\left[1-\lambda+\lambda e^{-2 i \Phi(t)} e^{s(t)}\right],
\end{equation}
where \cite{Dajka2008,Mierzejewski2009}

\begin{equation}\label{rsphi}
\begin{gathered}
r(t)=4 \int_0^{\infty} d \omega J^2(\omega)[1-\cos (\omega t)], \\
s(t)=  2 \int_0^{\infty} d \omega J(\omega) f(\omega)[1-\cos (\omega t)] -\frac{1}{2} \int_0^{\infty} d \omega f^2(\omega),\\
\Phi(t)=\int_0^{\infty} d \omega J(\omega) f(\omega) \sin (\omega t).
\end{gathered}
\end{equation}
In the above equations,  $f(\omega)$ and $J(\omega)$ are the real function and the effective spectral density of the environment respectively, which are given by
\begin{equation}
\begin{gathered}
f(\omega)=\omega^{\frac{\nu-1}{2}} \exp(-\omega/2 \omega_c),\\
J(\omega)=\sqrt{\alpha}\omega^{\frac{\mu-1}{2}}\exp(-\omega/2 \omega_c),
\end{gathered}
\end{equation}
where $\alpha$ is positive constant that describes the system-environment coupling, $\omega_c$ is the cutoff frequency, $\mu$ is the ohmicity parameter. Notice, the three cases $-1<\mu<0$, $\mu=0$, and $\mu>0$ correspond to the sub-ohmic, ohmic, and super-ohmic environments, respectively \cite{Zhang2015fan}. To study the evolution of the system in the model, the dynamics can be examined in two cases: ($i$) sub-ohmic and ohmic environments and ($ii$) super-ohmic environment. From a fundamental point of view,  it is preferable to consider the super-ohmic case $\mu>0$. From \eqref{rsphi}, one can obtain the following equations for the super-ohmic environment
\begin{equation}
\begin{gathered}
r(t)=4 \omega_c^\mu \alpha \Gamma(\mu) \left\{1-\frac{\cos \left[\mu \arctan \left(\omega_c t\right)\right]}{\left(1+\omega_c^2 t^2\right)^{\mu / 2}}\right\}, \\
s(t)=   2 \omega_c^{\chi} \sqrt{\alpha}\Gamma(\chi)\left\{1-\frac{\cos \left[\chi \arctan \left(\omega_c t\right)\right]}{\left(1+\omega_c^2 t^2\right)^{\chi / 2}}\right\}
- \frac{\omega_c^v}{2}\Gamma(v), \\
\Phi(t)=\omega_c^{\chi} \sqrt{\alpha} \Gamma(\chi)  \frac{\sin \left[\chi \arctan \left(\omega_c t\right)\right]}{\left(1+\omega_c^2 t^2\right)^{\chi / 2}},
\end{gathered}
\end{equation}
where $\chi=\frac{\mu+v}{2}$, and $\Gamma(.)$ is the Euler gamma function.

\section{Non-Markovianity}\label{Non-Markovian}
According to the structural features of the environment, the quantum evolution can be classified into two categories: (a) Markovian (without memory) and (b) non-Markovian (with memory). In the Markovian process, the environment acts as a waster for the system information and information sinks from the system into the environment. Actually, in Markovian evolution, information leaks from the system to the environment and there is no back-flow from the environment to the system. While there exists a back-flow of information from the environment to the system during the evolution in the non-Markovian case. Up to now, several computational criteria  have been introduced for the qualitative study of the non-Markovianity of quantum evolution \cite{non1,non2,non3,non4,non5,non6,non7,non8,non9,non10,non11,non12,non13,non14,non15,non16}
. In Ref. \cite{non2}, Breuer et al. have used state distinguishability to quantify the degree of non-Markovianity. They have interpreted the increment of state distinguishability as the return of information from the environment to the system. In this work, based on the states distinguishability, the non-Markovian criterion is considered as follows 
\begin{equation}\label{nonmarkovianity}
N=\max _{\rho_{1,2}(0)} \int_{\sigma>0} d t \sigma\left(t\right),
\end{equation}
In the above relation, $ \sigma\left(t\right)$ is the time derivative of trace distance
\begin{equation}
\sigma\left(t\right)=\frac{d}{d t} D\left(\rho_1(t), \rho_2(t)\right),
\end{equation}
where $D\left(\rho_1, \rho_2\right)=\frac{1}{2}\operatorname{tr}\left|\rho_1-\rho_2\right|$ is the trace distance that quantifies the distinguishability between two quantum states $\rho_1$ and $\rho_2$ (note that $\left|A\right|=\sqrt{A^\dag A}$ and $0 \leq D \leq 1$). It should be noted that for the whole dynamical semigroups and all time-dependent Markovian evolution, we have $\sigma(t) \leq 0$ while for $\sigma(t)>0$, the evolution is non-Markovian. In other words, it can be said that in non-Markovian evolution, distinguishability increases in some time intervals. From Eq. (\ref{nonmarkovianity}), it is clear that quantifying the degree of non-Markovianity needs to perform an optimization process over all pairs of initial states $\rho_{1,2}(0)$.

In Ref. \cite{non17}, it has been shown that the optimal state pairs are orthogonal. Therefore,  orthogonal states $\rho^{\lambda=0}_1(0)=\vert + \rangle \langle + \vert$ and $\rho^{\lambda=0}_2(0)=\vert - \rangle \langle - \vert$, where $\vert \pm \rangle =  \left( \vert e \rangle \pm \vert g \rangle \right)/\sqrt{2}$, can be considered as optimal states. For these optimal states, the trace distance at time $t$ is obtained as
\begin{equation}
D(\rho_1(t),\rho_2(t))= \vert \kappa_{\lambda=0}(t) \vert^{2}.
\end{equation}
where $\kappa_{\lambda}(t)$ is presented in Eq. \eqref{kappat}.

\begin{figure*}[ht]
\centering
    \includegraphics[width=0.5\linewidth]{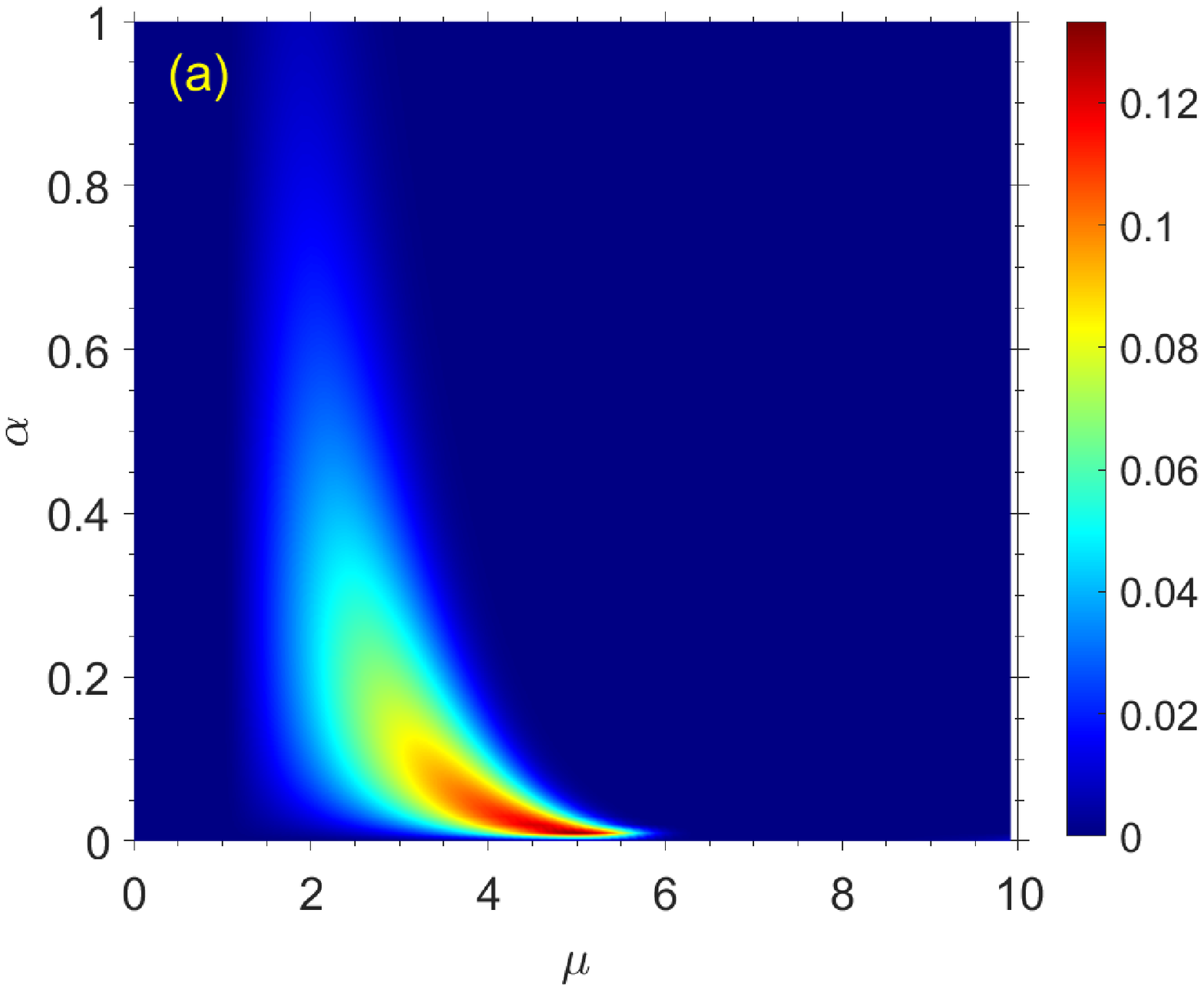}\hfil
    \includegraphics[width=0.5\linewidth]{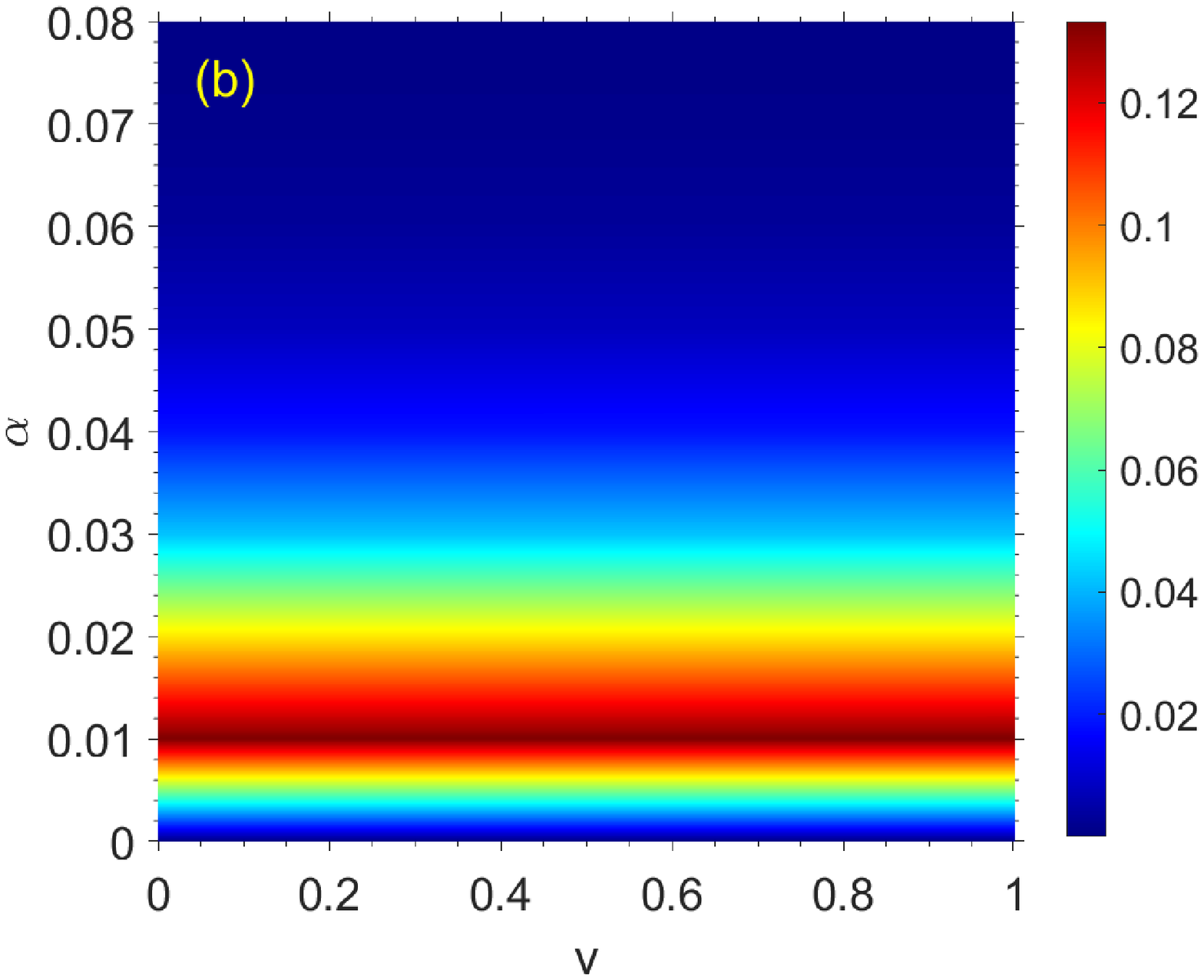}\par\medskip
    \centering
\caption{(a) Non-Markovianity $N$ as a function of coupling constant $\alpha$ and ohmicity parameter $\mu$ with  $v=0.01$. (b)  $N$ as a function of  $\alpha$ and $v$ with $\mu=5$.  For two plots $\lambda=0$ and $\omega_c=\omega_0=1$.}\label{1}
\end{figure*}

Now,  the effects of the ohmicity parameter $\mu$ and coupling constant $\alpha$ on the degree of non-Markovianity $N$ are investigated. Figure \ref{1} shows the degree of non-Markovianity as functions of $\mu$,  $\alpha$, and $v$ with fixed values $\lambda=0$ and $\omega_c=\omega_0=1$. In Fig. \ref{1}(a),  non-Markovianity is represented as functions of  $\mu$ and  $\alpha$ with $v=0.01$. As can be seen from this plot, for $\alpha\simeq 0.01$, the degree of non-Markovianity has its maximum value at $\mu \simeq 5$, but for $\mu>6$ and $\mu < 1$, it is equal to zero for all values of $\alpha$, indicating the Markovian dynamics. It can also be noticed that for different values of ohmicity parameter in a certain interval $1\leq \mu \leq 6$, the degree of non-Markovianity decreases by increasing $\alpha$.

Fig. \ref{1}(b) represents the non-Markovianity as  functions of $\alpha$ and $v$ with $\mu=5$. We see that the degree of non-Markovianity has its maximum value for $\alpha=0.01$. It can also be detected that for all values of $v$, the degree of non-Markovianity decreases when the value of $\alpha$ decreases or increases from $\alpha=0.01$. By comparing Fig. \ref{1}(a) and Fig. \ref{1}(b), one can conclude that to have a non-Markovian evolution for the considered model, it is enough to hold $\mu=5$ and $\alpha=0.01$, regardless of the value of $v$. In a similar way, it can be said that for $\mu>6$, the evolution is Markovian regardless of the values of $\alpha$ and $v$.

\section{Quantum Speed limit}\label{QSL}
In this section, the QSL is investigated for our considered model. Due to the initial correlation between the system and the environment, it is not possible for the initial state of the system to be pure. So, to calculate the QSL, it should be considered that the initial state is mixed. In Refs. \cite{Zhang2014,xiong2018}, the authors have introduced QSL for the mixed initial states. Here, the method presented in Ref. \cite{xiong2018} is used to study the QSL for the considered model.

The evolution of open quantum systems can be described by
\begin{equation}
\dot{\rho}_t=\mathcal{L}_t (\rho_t),
\end{equation}
in which $\mathcal{L}_t$ denotes the time-dependent positive generator. In Ref. \cite{xiong2018}, the authors used the function of relative purity \cite{Campaioli2018} as a distance measure, which is given by
\begin{equation}
\Theta\left(\rho_0, \rho_t\right)=\arccos \left(\sqrt{\frac{\operatorname{tr}\left[\rho_0 \rho_t\right]}{\operatorname{tr}\left[\rho_0^2\right]}}\right).
\end{equation}
Based on above function, the ML-type QSL bound for open quantum systems can be obtained as \cite{xiong2018}
 \begin{equation}\label{ML2}
\tau_{QSL}^{ML}=\max \left\{\frac{1}{\Lambda_\tau^{\mathrm{op}}}, \frac{1}{\Lambda_\tau^{\mathrm{tr}}}\right\} \sin ^2\left[\Theta\left(\rho_0, \rho_\tau\right)\right] \operatorname{tr}\left[\rho_0^2\right],
\end{equation}
with $\Lambda_\tau^{\mathrm{op(tr)}}=\frac{1}{\tau} \int_0^\tau d t\left\|\mathcal{L}_t\left(\rho_t\right)\right\|_{\mathrm{op(tr)}}$, where  $\left\|\mathcal{L}_t\left(\rho_t\right)\right\|_{\mathrm{tr}}=\sum_i \lambda_i$ and $\left\|\mathcal{L}_t\left(\rho_t\right)\right\|_{\text {op }}=\lambda_1$ are  the trace norm and operator norm  for  $\mathcal{L}_t\left(\rho_t\right)$. Herein, $\lambda_i$'s and $\lambda_1$ are singular values and largest singular value of  $\mathcal{L}_t\left(\rho_t\right)$.

Also, the MT-type bound on the QSL for non-unitary dynamics can be expressed as follows
\begin{equation}\label{MT2}
\tau_{QSL}^{MT}=\frac{1}{\Lambda_\tau^{\mathrm{hs}}} \sin ^2\left[\Theta\left(\rho_0, \rho_\tau\right)\right] \operatorname{tr}\left[\rho_0^2\right],
\end{equation}
with $\Lambda_\tau^{\mathrm{hs}}=\frac{1}{\tau} \int_0^\tau d t\left\|\mathcal{L}_t\left(\rho_t\right)\right\|_{\mathrm{hs}}$ where $\|\mathcal{L}_t\|_{\mathrm{hs}}=\sqrt{\sum_i \lambda_i^2}$ is the Hilbert-Schmidth norm of  $\mathcal{L}_t\left(\rho_t\right)$. Combining Eqs. (\ref{ML2}) and (\ref{MT2}), the unified bound on the QSL for non-unitary dynamics can be formulated as
 \begin{equation}\label{QSLc}
\tau_{QSL}:=\max \left\{\frac{1}{\Lambda_\tau^{\mathrm{op}}}, \frac{1}{\Lambda_\tau^{\mathrm{tr}}}, \frac{1}{\Lambda_\tau^{\mathrm{hs}}}\right\} \sin ^2\left[\Theta\left(\rho_0, \rho_\tau\right)\right] \operatorname{tr}\left[\rho_0^2\right].
\end{equation}
Notice that the ML-type bound based on the operator norm is the sharpest QSL bound for non-unitary dynamics.

Now, the QSL can be checked for the considered model. By putting $t=0$ and $c_{e,g}=1/\sqrt{2}$ in Eq. (\ref{rhot}), the correlated initial  state is obtained as
\begin{equation}\label{rho0}
\rho_0=\frac{1}{2}\left(\begin{array}{ll}
1 &  \kappa_\lambda(0) \\
 \kappa_\lambda^*(0) & 1
\end{array}\right).
 \end{equation}
 Hence, from Eq. (\ref{QSLc}), the QSL for the above correlated initial state can be obtained as
 \begin{equation}\label{QSLCd1}
 \tau_{QSL}=\frac{\vert \kappa_\lambda(0) \vert ^{2}-\vert \kappa_\lambda(0) \vert \operatorname{Re}[\kappa_\lambda(\tau)]}{\int_0^{\tau}\vert \dot{\kappa}_\lambda(t)\vert dt},
 \end{equation}
 where $\tau$ is actual driving time.

 In order to show the effect of initial quantum coherence of the correlated initial state on the QSL, it is necessary to consider an analytic quantifier of quantum coherence \cite{coh1,coh2,coh3,coh4,coh5}. Here, the $l_1$-norm of coherence is considered to quantify the quantum coherence as $C(\rho)=\sum_{i\neq j} \vert \rho_{ij} \vert$. The $l_1$-norm  quantum coherence of correlated initial state (\ref{rho0}) can be obtained as $C(\rho_0)=\vert \kappa_\lambda(0) \vert$. So, it can be seen that the initial quantum coherence depends on the initial correlation between the system and the environment, i.e., $\lambda$. The quantum coherence has its maximum value one for  uncorrelated case $\lambda=0$ and it is equal to zero for  fully correlated case  $\lambda=1$. Thus, the QSL in Eq. (\ref{QSLCd1}) can be rewritten as
 \begin{equation}\label{QSLCd}
 \tau_{QSL}=\frac{C(\rho_0) ^{2}- C(\rho_0)  \operatorname{Re}[\kappa_\lambda(\tau)]}{\int_0^{\tau}\vert \dot{\kappa}_\lambda(t)\vert dt}.
 \end{equation}
 \begin{figure}
\centerline{\includegraphics[width=9cm]{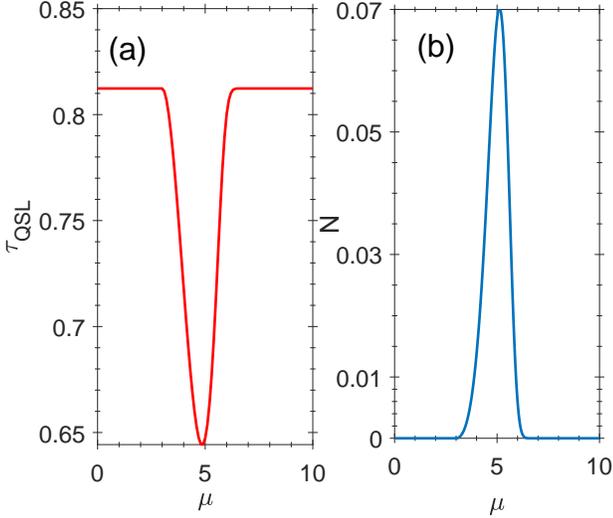}}
\caption{(a) QSL for a correlated initial state \eqref{rho0} as a function of ohmicity parameter $\mu$  with  $\lambda=0.25$  and $\tau=1$. (b) Non-Markovianity in terms of $\mu$ with $\lambda=0$. For two plots $\alpha=v=0.01$, and $\omega_c=\omega_0=1$.}\label{2}
\end{figure}
Fig. \ref{2} shows the changes of both QSL and non-Markovianity  in terms of the ohmicity  parameter $\mu$. In Fig. \ref{2}(a), the QSL is sketched as a function of $\mu$ for correlated initial state $\lambda=0.25$ with $v=\alpha=0.01$ and driving time $\tau=1$. Besides, Fig.\ref{2}(b) represents the non-Markovianity in terms of $\mu$ with $\lambda=0$ and $\alpha=v=0.01$. Comparing plots \ref{2}(a) and \ref{2}(b),  we find that there exists an inverse qualitative relationship between the QSL  and non-Markovianity. Notably, for $\mu=5$, the degree of non-Markovianity has its maximum value, while the QSL has its minimum value (see also Fig. \ref{1}).
\begin{figure}
\centerline{\includegraphics[width=9cm]{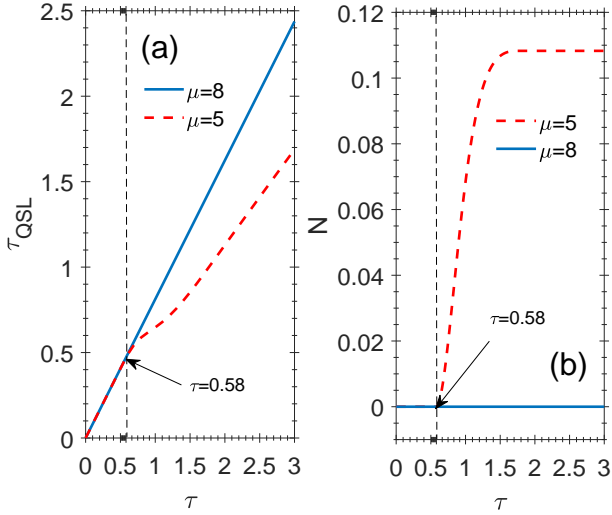}}
\caption{(a) QSL time as a function of driving time $\tau$ for both Markovian $\mu=8$ and non-Markovian $\mu=5$ environments with $\lambda=0.25$. (b) Non-Markovianity versus $\tau$ for $\mu=8$ and $\mu=5$ with $\lambda=0$. For two plots $\alpha=v=0.01$, and $\omega_c=\omega_0=1$.}\label{3}
\end{figure}

In Fig. \ref{3}, the QSL and non-Markovianity are plotted as a function of driving time $\tau$. Fig. \ref{3}(a) shows the QSL for both Markovian $\mu=8$ and non-Markovian $\mu=5$ environments. As can be seen, for non-Markovian regime the QSL is shorter than Markovian regime. On the other hand, Fig. \ref{3}(b) illustrates the degree of non-Markovianity as a function of  $\tau$. As expected, the non-Markovianity is equal to zero for all driving time when $\mu=8$. While for $\mu=5$, the non-Markovian nature of the evolution is revealed from a specific driving time. An interesting result can be detected by comparing plots \ref{3}(a) and \ref{3}(b) is that before the appearance of the non-Markovian nature for $\mu=5$, the QSL for both $\mu=8$ and $\mu=5$ are coincide while, with the appearance of the non-Markovian nature at time $\tau=0.58$, the QSL becomes shorter than the case with $\mu=8$. Remarkably, it can be concluded that the non-Markovianity nature leads to the speedup of quantum evolution.
\begin{figure}
\centerline{\includegraphics[width=9cm]{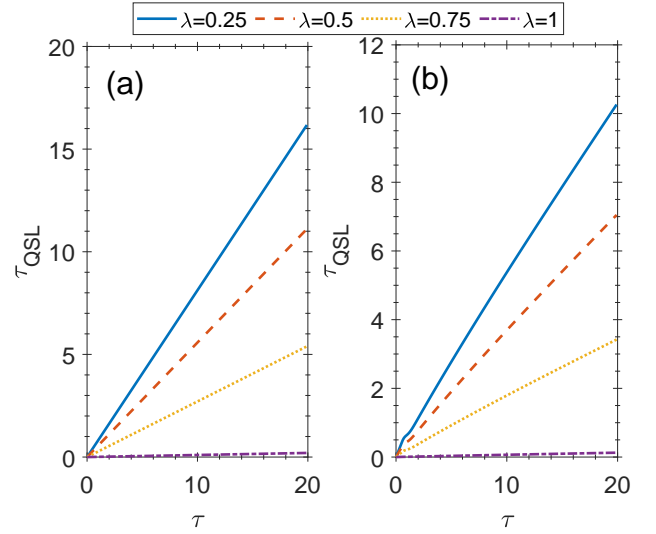}}
\caption{QSL as a function of driving time $\tau$ for different values of initial correlation between system and environment, $\lambda=0.25, 0.50, 0.75,$ and $1$. (a) Markovian regime $\mu=8$ and (b) non-Markovian regime $\mu=5$.  For two plots $\alpha=v=0.01$, and $\omega_c=\omega_0=1$. }\label{4}
\end{figure}

In Fig.\ref{4}, the QSL is illustrated as a function of driving time for different values of correlation parameter $\lambda=0.25, 0.50, 0.75,$ and $1$ in both Markovian and non-Markovian regimes. These plots reveal the effect of initial correlation on the QSL in Markovian and non-Markovian cases. It can be found that for both Markovian and non-Markovian cases, the QSL will be shorter with increasing the initial correlation between the system and the environment.
\begin{figure}
\centerline{\includegraphics[width=9cm]{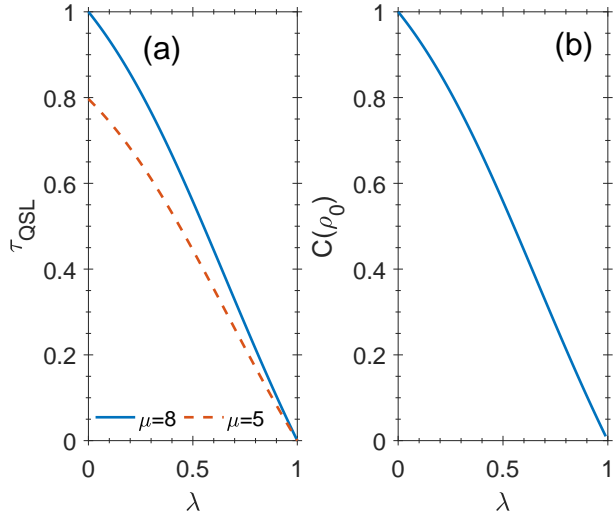}}
\caption{(a) QSL versus the correlation parameter $\lambda$ for Markovian $\mu=8$ and non-Markovian  $\mu=5$ regimes. (b) The $l_1$-norm of coherence $C(\rho_0)$ in terms of $\lambda$. For two plots $\alpha=v=0.01$, $\omega_c=\omega_0=1$, and  $\tau=1$.}\label{5}
\end{figure}

In Fig. \ref{5}(a), the QSL is plotted as a function of correlation parameter $\lambda$ with driving time $\tau=1$. From  Fig. \ref{5}(a), for both Markovian $\mu=8$ and non-Markovian $\mu=5$ evolutions, the QSL becomes shorter as the correlation parameter increases. As shown in Eq. (\ref{QSLCd}), the QSL depends on the coherence of the initial state of the system. Therefore, to justify the result obtained from Fig. \ref{5}(a), the coherence of the initial state of the system is drawn in terms of  $\lambda$ in Fig. \ref{5}(b). As expected, the quantum coherence of the initial state of the system diminishes with increasing the correlation parameter $\lambda$.
\begin{figure}
\centerline{\includegraphics[width=9cm]{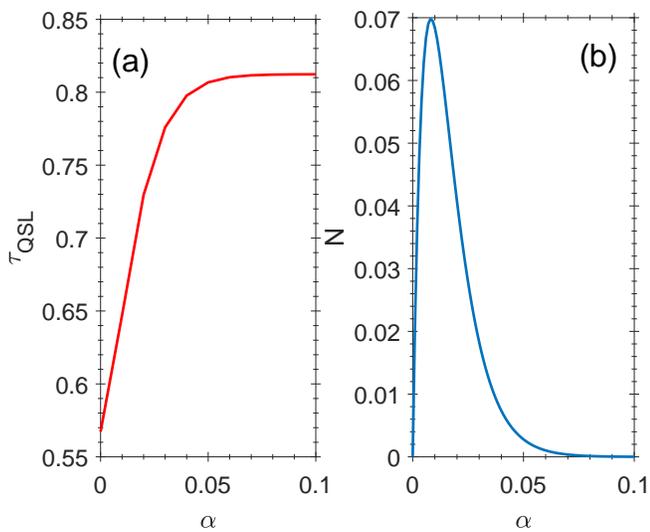}}
\caption{(a) QSL as a function of coupling parameter $\alpha$ for driving time $\tau=1$ with $\lambda=0.25$ and (b) non-Markovianity versus $\alpha$ with $\lambda=0$. For two plots $v=0.01$, $\omega_c=\omega_0=1$, and  $\mu=5$. }\label{6}
\end{figure}

In Fig. \ref{6}(a) the QSL is sketched as a function of coupling parameter $\alpha$. As can be seen, the QSL grows with the increase of the coupling parameter $\alpha$ and reaches a constant value at $\alpha=0.1$.   Fig. \ref{6}(b) represents the non-Markovianity in terms of coupling parameter $\alpha$. Again, from plots \ref{6}(a) and \ref{6}(b) as well as Fig. \ref{1}, one can observe that in the non-Markovian regime ($\alpha\approx0.01$), the QSL is shorter than in the Markovian case ($\alpha\approx0.1$).

\begin{figure}
\centerline{\includegraphics[width=9cm]{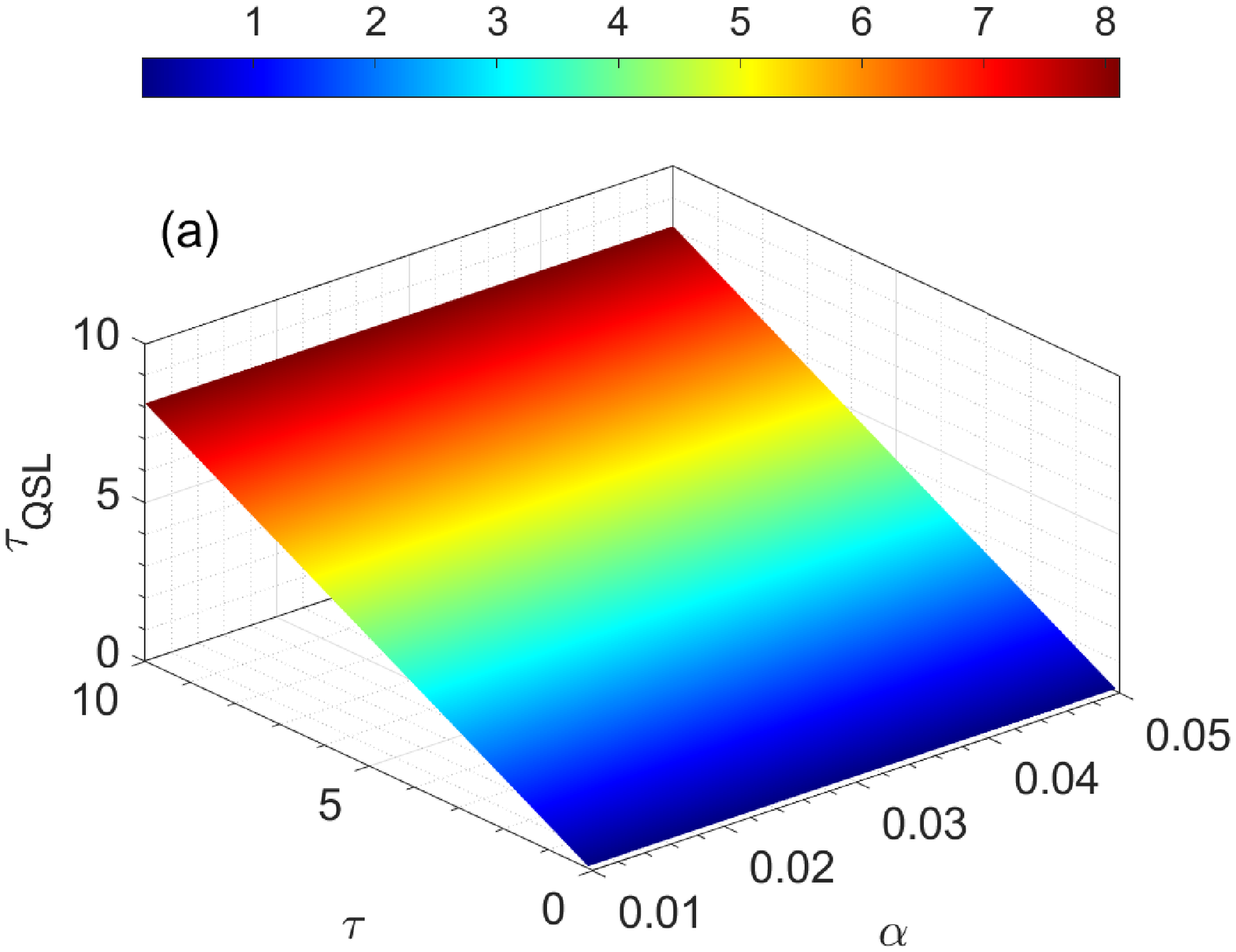}}
\centerline{\includegraphics[width=9cm]{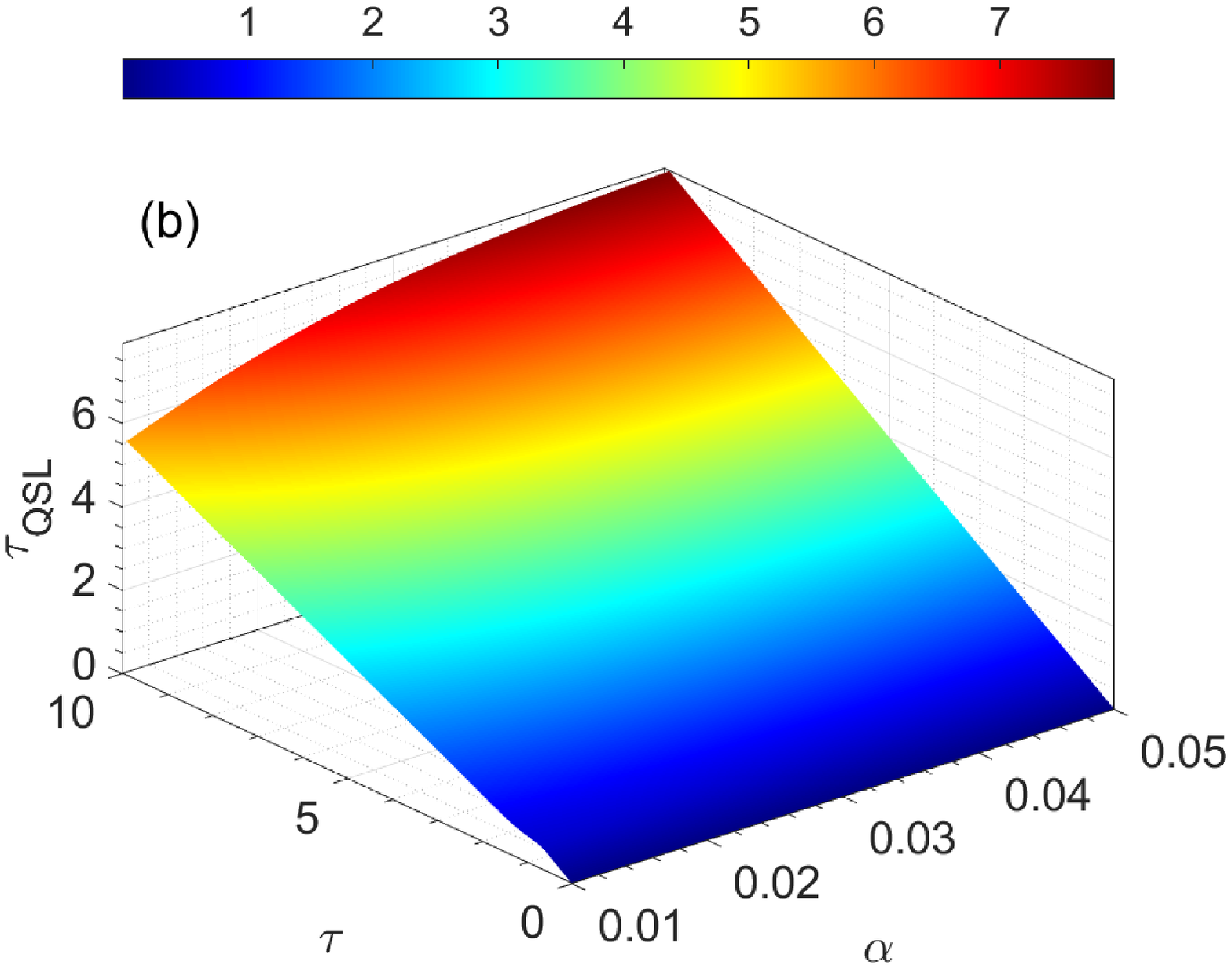}}
\caption{QSL as functions of coupling parameter $\alpha$ and driving time $\tau$ with $v=0.01$, $\omega_c=\omega_0=1$, and $\lambda=0.25$. (a) Markovian $\mu=8$ (b) non-Markovian $\mu=5$ regimes.  }\label{7}
\end{figure}
Finally, Fig. \ref{7} displays the QSL as functions of the coupling parameter $\alpha$ and driving time $\tau$ for both Markovian and non-Markovian dynamics when the correlation parameter is $\lambda=0.25$. From Fig. \ref{7}(a), we find that the QSL does not alter with $\alpha$ changes for the Markovian case. However, Fig. \ref{7}(b) shows that in a non-Markovian regime, the QSL increases as $\alpha$ growths.

\section{Conclusion}\label{conclusion}
The QSL has studied in an open quantum system with the initial correlation between the system and the environment. Specifically, we used the  QSL bound based on the function of the relative purity introduced in Ref. \cite{xiong2018}. First, we examined the considered model from the aspect of memory effects and determined the range of environmental parameters that cause the process to be non-Markovian. After that, we examined the effects of non-Markovianity of quantum evolution on QSL and found that the non-Markovian effects lead to shorter QSL. In other words, non-Markovian effects can speedup quantum evolution. We also observed that the initial coherence of the quantum system is directly related to the QSL, e.g., the highest QSL belongs to the initial  states with the higher quantum coherence. As another result of this work, we found that increasing the initial correlation between the system and the environment leads to a decrease in the QSL. Indeed, the effect of the initial correlation between the system and environment on QSL originates from the dependence of QSL on quantum coherence. Moreover, we revealed that the increasing coupling parameter leads to a boost in QSL for the non-Markovian evolution.


\section*{Data availability}
No datasets were generated or analyzed during the current study.

\section*{Competing interests}
The authors declare no competing interests.

\section*{ORCID iDs}
\noindent Maryam Hadipour\\ \href{https://orcid.org/0000-0002-6573-9960}{https://orcid.org/0000-0002-6573-9960}\\
Soroush Haseli\\ \href{https://orcid.org/0000-0003-1031-4815}{https://orcid.org/0000-0003-1031-4815}\\
Saeed Haddadi\\ \href{https://orcid.org/0000-0002-1596-0763}{https://orcid.org/0000-0002-1596-0763}\\
Hazhir Dolatkhah\\ \href{https://orcid.org/0000-0002-2411-8690}{https://orcid.org/0000-0002-2411-8690}


\end{document}